\documentclass[12pt]{article}
\usepackage{amsmath}
\usepackage{graphicx,psfrag,epsf}
\usepackage{enumerate}
\usepackage{natbib}
\usepackage{url} 

\usepackage{amsgen,amsthm,amsmath,amstext,amsfonts,amsbsy,amsopn,amssymb,latexsym}

\usepackage[usenames]{color}
\usepackage{array}
\usepackage{cite}
\usepackage{todonotes}
\usepackage{multirow}
\usepackage{graphicx}
\usepackage{amssymb}
\usepackage{natbib}
\usepackage{tikz}
\usepackage{xcolor}
\usetikzlibrary{positioning}
\usepackage{hyperref}
\usepackage{amsfonts}
\usepackage{algorithm}
\usepackage[noend]{algpseudocode}
\usepackage{subcaption}
\usepackage{amsfonts}
\usepackage{booktabs}
\usepackage{setspace}
\usepackage{textcomp}
\usepackage{scalerel}
\usepackage{accents}

\newcommand{\blind}{1}

\newtheorem{lemma}{Lemma}
\newtheorem{prop}{Proposition}

\newtheorem{coro}{Corollary}
\newtheorem{remark}{Remark}
\newtheorem{theorem}{Theorem}
\newtheorem{example}{Example}
\def\X{\boldsymbol{X}}
\def\Z{\boldsymbol{Z}}

\newcommand{\barJ}{\bar{\mkern-2mu\boldsymbol{J}\mkern2mu}}
\newcommand{\barS}{\bar{\mkern-2mu\boldsymbol{S}\mkern2mu}}

\def\S{\boldsymbol{S}}

\def\Lsc{\mathcal{L}}
\def\Msc{\mathcal{M}}
\def\Csc{\mathcal{C}}
\def\Usc{\mathcal{U}}
\def\trans{^{\top}}

\def\c{\boldsymbol{c}}
\def\Ehat{\widehat{{\rm E}}}

\def\bdelta{\boldsymbol{\delta}}

\def\var{{\rm Var}}
\def\varhat{\widehat{{\rm Var}}}

\def\Op{O_{\sf P}}

\def\A{\boldsymbol{A}}
\def\W{\boldsymbol{W}}
\def\w{\boldsymbol{w}}

\def\op{o_{\sf P}}

\newcommand{\citeg}[1]{\citep[e.g.,][]{#1}}
\def\Acal{\mathcal{A}}
\def\bphi{\boldsymbol{\phi}}
\def\tbphi{\widetilde{\boldsymbol{\phi}}}
\def\blambda{\boldsymbol{\lambda}}
\def\cF{\mathcal{F}}
\def\DEF{\scaleto{\rm DEF\mathstrut}{5pt}}

\allowdisplaybreaks[4]

\begin{document}

\def\spacingset#1{\renewcommand{\baselinestretch}%
{#1}\small\normalsize} \spacingset{1}
\newcommand\ryw[1]{{\color{teal}Ruoyu: ``#1''}}

\if1\blind
{
  \title{\bf Adaptive and Efficient Learning with Blockwise Missing and Semi-Supervised Data}
  \author{Yiming Li$^1$, Ruoyu Wang$^2$\footnote{Li and Wang contributed equally to this work.}, Ying Wei$^1$ and Molei Liu$^{3}$ \\
  \\~
{1. Department of Biostatistics, Columbia University} \\
{2. Department of Biostatistics, Harvard University} \\
{3. Department of Biostatistics, Peking University} \\
    }
    \date{}
  \maketitle
} \fi

\if0\blind
{
  \bigskip
  \bigskip
  \bigskip
  \begin{center}
    {\LARGE\bf Adaptive Learning with Blockwise Missing and Semi-Supervised Data}
\end{center}
  \medskip
} \fi

\begin{abstract}

\noindent Data fusion enables powerful and generalizable analyses across multiple sources. However, different data collection capacities across different sources lead to blockwise missingness (BM), which poses challenges in practice. Meanwhile, the high cost of obtaining gold-standard labels leaves the majority of samples unlabeled, known as the semi-supervised (SS) problem. In this paper, we propose a novel {\bf D}ata-adaptive {\bf E}stimation approach for data {\bf FU}sion in the {\bf SE}mi-supervised setting (DEFUSE) that handles both BM and SS issues in the presence of distributional shifts across data sources under a conditional alignment assumption. DEFUSE starts with a complete-data-only estimator derived from the primary data source, and uses data-adaptive and {distributional-shift-adjusted} procedures to successively incorporate the data with BM covariates and the large unlabeled sample to effectively reduce the estimation variance without incurring bias. {To further avoid bias due to fusion of misaligned data violating of the transportability assumption, a screening method is developed to identify and exclude data sources that are not aligned with the primary source.} DEFUSE offers a new data-adaptive control variate approach to handle BM, which achieves intrinsic efficiency and robustness against distributional shifts. {Furthermore, DEFUSE can attain the semiparametric efficiency bound when the control variate is properly specified.} These advantages are theoretically guaranteed and empirically supported by simulation studies and two real-world biomedical applications. 

\end{abstract}

\noindent%

\spacingset{1.85}  

\section{Introduction}


\subsection{Background}
Data fusion is the process of integrating multiple data sources to obtain richer and more comprehensive insights than what could be achieved by analyzing each source separately. It has been increasingly used in many applications.  For example, in the field of health and medicine, there has been growing interest and effort in linking electronic medical records (EMRs) with biobank data to tackle complex health issues \citeg{castro2022mass}. EMRs include detailed longitudinal clinical observations of patients, offering a rich source of health information over time. By combining these records with multi-omics data from biobanks, this type of data fusion allows for a much deeper understanding of disease prognosis at the molecular level, potentially transforming patient care and treatment strategies \citeg{zengini2018genome}. Beyond integrating different types of data, data fusion has also been scaled up to merge data from multiple institutions. The All-of-Us Research Program \citep{all2019all} is a prime example.

To effectively analyze fused datasets, particularly those combined with EMR, several major statistical challenges must be addressed. {\em Distributional shift} across data sources has been frequently studied in the past years \citeg{gretton2009covariate}. Beyond this challenge, structural heterogeneity of multi-source data sets can severely impede data fusion. One specific common issue is known as \textit{blockwise missingness} (BM) of covariates, which arises when variables are collected or defined differently across sources, resulting in one or multiple blocks of missing data  in the merged data. Additionally, EMR-related research often suffers from incomplete or unreliable outcome labels.  Obtaining accurate labels  requires extensive manual review by clinical experts or long-term  follow-up with patients. That leaves many patient outcomes unobserved/unlabeled in EMR. In such cases,
\textit{semi-supervised} (SS) learning is particularly valuable, as it combines a small subset of accurately labeled data with a large unlabeled sample from the same population. In situations where EMR data is linked with other data resources, the issues of BM and SS  often occur simultaneously; see the real-world case study in Section \ref{sec:real}. This new setup involves a more complex data structure, which we formally describe in Section \ref{sec:intro:problem}. The complexity of EMR-fused data motivates our methodological development. 


\subsection{Problem Setup}\label{sec:intro:problem}

Let $Y$ be the outcome of interest and let $\X = (X_1, X_2, \ldots, X_p)^\top$ denote a $p$-dimensional vector of covariates. 
There are three types of data structures in a fused data. (i) \textbf{Labeled Complete ($\Lsc\Csc$)} data source, which consists of labeled and complete observations. (ii) \textbf{Labeled Missing sources ($\Lsc\Msc_r : r = 1, \ldots, R$)} with labeled outcomes but parts of covariates are missing in blocks. In data source $\Lsc\Msc_r$, only $(Y, \X_{\Gamma_r})$ are observed, where $\Gamma_r \subset \{1, 2, \ldots, p\}$ is the index set of the observed covariates, and $\X_\mathcal{B}$ denotes the subvector of $\X$ containing $X_j$ for $j \in \mathcal{B}$. Let $\Gamma_r^c = \{1,2,\ldots,p\} \setminus \Gamma_r$ be the complement of $\Gamma_r$. For example, if we observe the first two covariates $(X_1, X_2)$ and $p = 5$, then $\Gamma_r = \{1, 2\}$, $\Gamma_r^c = \{3, 4, 5\}$, $\X_{\Gamma_r} = (X_1, X_2)^\top$, and $\X_{\Gamma_r^c} = (X_3, X_4, X_5)^\top$;  (iii) \textbf{Unlabeled Complete ($\Usc\Csc$)} data with unobserved outcomes but fully observed covariates.

The complicated non-monotone BM data structure is encountered frequently in real world. For example, $\Lsc\Csc$ and $\Lsc\Msc_r$'s are collected from multiple biomedical studies with the outcome $Y$ costly to obtain in some larger observational cohort $\Usc\Csc$.
We denote $n$, $n_r$ and  $N$ as the sample sizes of $\Lsc\Csc$, $\Lsc\Msc_r$ and $\Usc\Csc$, respectively. We use the source label as the subscript of the expectation operator $\rm E$ to indicate the population with respect to which the expectation is taken. In the semi-supervised (SS) setting, we assume that $N \gg n + \sum^R_{r = 1}n_r$ and $R$ is fixed. We also define $\rho_r = n_r / n$. 
Our primary goal is to infer a generalized linear model (GLM) on the $\Usc\Csc$ population, with the parameter of interest $\bar{\boldsymbol{\gamma}}$ defined as the solution to the population-level equation:
\begin{equation}
\mathbf{U}(\boldsymbol{\gamma}) = {\rm E}_{\Usc\Csc}\left[\A \left\{Y - g(\boldsymbol{\gamma}^\top \A)\right\}\right] = \mathbf{0},
\label{eq:ee:pop}
\end{equation}
where $g(\cdot)$ is a known link function with derivative $\dot{g}$ and $\A$ is a subset of the covariates in $\X$ of our interest in terms of risk prediction. For example, $g(a) = a$ corresponds to a Gaussian linear model, and $g(a) = e^a / (1 + e^a)$ for the logistic model. When the GLM is correctly specified, i.e., $\mathrm{E}_{\Usc\Csc}[Y\mid \A] = g(\boldsymbol{\gamma}^{*\top} \A)$ for some $\boldsymbol{\gamma}^*$, $\bar{\boldsymbol{\gamma}}$ is identical to the true model parameter $\boldsymbol{\gamma}^*$. When the target model is misspecified, $\bar{\boldsymbol{\gamma}}$ can be viewed as a pseudo-true value that solves the equation \eqref{eq:ee:pop} and is still useful in practice, e.g., the clinical risk score or polygenic risk score for certain disease \citeg{elliott2020predictive}. Our method and theory hold regardless of whether this model is correctly specified or not. 

In this paper, our primary goal is to leverage and fuse multiple data sources for robust and efficient inference of $\bar{\boldsymbol{\gamma}}$ defined on the source $\Usc\Csc$.
Importantly, the distributional heterogeneity of variables between sources is almost inevitable due to the varying measurement standards or patient recruitment criteria. Under such heterogeneity, the usual Missing Completely at Random (MCAR) assumption, which is commonly used in BM data literature, becomes unrealistic, as missingness often depends on shared or study-specific covariates. To address this, we incorporate possible distributional shifts across data sources and allow missingness to depend on observed variables, consequently relaxing MCAR to a more flexible partial alignment assumption. This enables more realistic and generalizable data fusion, while requiring careful handling of distributional shifts.

Let $P_{\Lsc\Csc}$, $P_{\Lsc\Msc_r}$ and $P_{\Usc\Csc}$ denote the underlying distribution of the corresponding data sources, and $p_{\Lsc\Csc}$, $p_{\Lsc\Msc_r}$ and $p_{\Usc\Csc}$ denote the corresponding densities. We consider the scenario with distributional shifts across the data sources. In particular, the target population $\Usc\Csc$ can have a different distribution as $\Lsc\Csc$ and $\Lsc\Msc_r$, i.e., $P_{\Usc\Csc}\neq P_{\Lsc\Csc}$ and/or $P_{\Usc\Csc}\neq P_{\Lsc\Msc_r}$. To formally address this, we explicitly distinguish between two related but different components of our problem. {The first is the covariate missing mechanism: within each $\mathcal{LM}_r$ block, $Y$ is assumed to be observed, while components of $\X$ may be missing and the missingness pattern of $\X$ is allowed to depend on observed information, including $Y$, observed components of $\X$, or observed auxiliary variables. This mechanism specifically concerns why certain covariates are unobserved within a source. The second component is the source-alignment or transportability condition required for integrating information across data sources, which dictates when information from $\mathcal{LM}_r$ can be validly transported to the target population $\Usc\Csc$. We assume this latter condition through the following Conditional Alignment under Missingness (CAM):
\begin{equation}\label{eq:cs:assume}
P_{\Lsc\Csc}(Y\mid\X)=P_{\Usc\Csc}(Y\mid\X),\quad P_{\Lsc\Msc_r}(Y,\X\mid\Z_r)=P_{\Usc\Csc}(Y,\X\mid\Z_r),     
\end{equation}
where $\Z_r$ consists of a prespecified subset of variables in $(\X_{\Gamma_r},Y)$. Crucially, a variable that affects the covariate-observation pattern does not necessarily need to be included in $\Z_r$; it needs to be included only when conditioning on it is necessary for source alignment. Under (\ref{eq:cs:assume}), the data distributions conditional on $\Z_r$'s are aligned and transferable between the $\Lsc\Msc_r$ and $\Usc\Csc$ samples. The assumption for $\Lsc\Csc$ in (\ref{eq:cs:assume}) is the commonly used covariate shift assumption \citeg{gretton2009covariate,liu2023augmented}.

Compared to conventional missing mechanisms including missing completely at random (MCAR) and missing at random (MAR), CAM is more appropriate for our data-fusion setting that aims at transferring knowledge from the labeled sources to the target unlabeled population, rather than the full-data population typically targeted in standard missing data analyses. In this context, CAM operates as a cross-source transportability condition mapping the $\Lsc\Msc$ sources directly to the target $\Usc\Csc$. Note that the MCAR assumption implies \eqref{eq:cs:assume} holds with $\Z_r = \emptyset$ and hence is a special case of CAM. Also, in the standard MAR setting, one usually does not include $\Usc\Csc$ and assumes that $P_{\Lsc\Msc_r}(Y,\X\mid Y, \X_{\Gamma_r}) = P_{F}(Y,\X\mid Y, \X_{\Gamma_r})$ with $F=\Lsc\Csc\cup\{\Lsc\Msc_r:r=1,\ldots,R\}$ and $P_{F}$ being the full-data distribution. This assumption neither implies nor is implied by CAM. Nevertheless, our CAM assumption is similar to MAR in the sense that it allows the blockwise missingness of covariates to depend on the observed $Y$, not just a subset of $X_{\Gamma_{r}}$.

Conceptually, CAM resembles the transportability assumption in the data fusion literature \citep{pearl2014external,Li2023efficient,qiu2024efficient}.
It enables valid knowledge transfer from data sources that are conditionally aligned on \(\Z_r\). 
One can refer to \citet{pearl2014external} and others for some principles to specify $\Z_r$ based on causal diagrams. However, prior knowledge may not be fully correct and certain sources of BM data might violate the alignment assumption in (\ref{eq:cs:assume}), with distributions of \((Y, \X_{\Gamma_r})\mid\Z_r\) different from the target population \(\mathcal{UC}\). We refer to such BM sources as \textit{misaligned}. Therefore, identifying and excluding misaligned sources becomes necessary to preserve the validity of our CAM-based inference. This consideration naturally motivates the development of a screening procedure to detect and remove such bias-prone sources before fusion as will be introduced in Section \ref{sec:violation}.}

\begin{remark}
We allow the inclusion of auxiliary or surrogate features in possibly high-dimensional $\Z_r$ or $\X_{\Gamma_r}$ that are informative about $Y$ or missing covariates but not necessarily included in the target model predictors $\A$. This is particularly relevant in EMR applications where a wealth of auxiliary information is routinely collected. For example, in building a risk prediction model for heart disease, the target covariates $\A$ might include biomarkers that are clinically interpretable and relevant to the disease mechanism. Meanwhile, $\X$ could contain auxiliary features such as medication codes and clinical mentions strongly correlated with the missing covariates and outcome; see Section \ref{sec:real:mimic}. These auxiliary variables can help aligning the data distributions across the sources or improve the imputation of missing data and enhance estimation efficiency, even though they are excluded from the final scientific model. This distinction between $\A$ and $\X$ allows practitioners to leverage the predictive power of high-dimensional or AI-generated auxiliary features \citeg{angelopoulos2023prediction}. 
\label{rem:intro}
\end{remark}

\subsection{Related Literature}\label{sec:intro:lit}

As more data is combined from different institutions, blockwise missing covariates have become an important problem. Several recent works have been designed for BM data to achieve more robust and efficient estimation. \citet{yu2020optimal} proposed an approach for integrative linear regression under BM. \citet{xue2021integrating} proposed to aggregate linear estimating equations with multi-source imputed data sets to achieve an efficiency gain. \citet{xue2021statistical} further extended this framework for high-dimensional inference. \citet{song2024semi} applied the idea of debiasing to handle BM without imputations, and \citet{jin2023modular} proposed a modular regression approach for multi-modal covariates. These models focused on the linear model only. For GLM allowing nonlinear links, \citep{li2023penalized} proposed a penalized estimating equation approach with multiple imputation. \citet{zhao2023heterogeneous} introduced a generalized method of moments (GMM) approach to incorporate the BM data. Their methods fit a GLM to the BM data using available covariates only and then using this ``reduced'' working model to improve the estimation of the target model. Similar strategies were also used in \citet{song2024semi} and others with different formulations. For example, \citet{wang2023survey} derived calibration weights based on the moment condition of the BM's reduced model, and used them to reweight the complete data to improve estimation efficiency. This idea dates back to the model-calibration method for survey data with auxiliary information \citep{wu2001model}.  

The pursuit of semiparametric efficiency has been a central theme in missing data literature. The seminal work of \citet{robins1994estimation} established the theory of efficient estimation with missing covariates. Recent advances have further addressed the BM problems. For instance, \citet{Li2023efficient} developed the efficient influence function that can be applied for fusing longitudinal data with monotone missing structures. \citet{qiu2024efficient} considered a similar setup as \citet{Li2023efficient} and proposed multiply robust estimators for risk estimation under general forms of dataset shift. {\citet{berrett2024efficient} focused on mean estimation of a known function under more general and challenging non-monotone block-missingness settings. \citet{graham2025unified} studied a general data fusion problem that allows data from multiple sources to exhibit non-monotone block-missingness structures. Their work represents the state of the art in terms of theoretical efficiency. However, its framework does not directly yield explicit influence-function representations, nor does it readily lead to estimators that are straightforward to implement. In practice, the efficiency gains based on their results may remain sensitive to nuisance-model misspecification. }

The two lines of work reviewed above do not explicitly address the practical issue of source misalignment, where distributional differences between data sources may violate the alignment assumptions like (\ref{eq:cs:assume}) required for valid inference. 
Recent work \citeg{cheng2021adaptive,han2025federated} addressed this problem through sensitivity analyses that leverage the difference of certain summary statistics to examine the alignment between sources and exclude misaligned ones. \citet{yang2023elastic} and \citet{chen2025minimax} considered a similar adaptive fusion strategy when combining experimental and observational studies for inference of treatment effects.

Our work also aligns with the recent progress in semi-supervised learning, which leverage large unlabeled datasets to enhance statistical efficiency when labeled data are limited. For example,  \citep{kawakita2013semi} and \citep{song2023general} proposed to reweight the regression on the labeled sample with estimated density ratios. \citep{chakrabortty2018efficient} and \citep{gronsbell2022efficient} imputed its unobserved outcome $Y$. \citep{wu2023robust} extended this imputation framework for SSL of graphic models. These SSL methods can produce more efficient estimates than their supervised learning (SL) counterpart when the outcome model is misspecified while they remain as efficient as SL when that model is correct. {Recently, \citet{liu2023augmented}, \citet{zhang2023double} and others extended SSL to handle covariate shifts through doubly robust estimation, but their framework does not accommodate the complex blockwise missing structures prevalent in fused data sources.}

\subsection{Our contribution}

We propose a novel \textbf{D}ata-adaptive \textbf{E}stimation approach for data \textbf{F}Usion in the \textbf{SE}mi-supervised setting (DEFUSE). It systematically handles both blockwise missingness and semi-supervised problems {in the CAM setting} through two key methodological innovations. They are supported by asymptotic analysis, simulations, and real-world applications.

We introduce a novel data-adaptive approach for fusing BM data under CAM. Our framework employs a family of control functions that maintain valid statistical inference while achieving efficiency gains. Compared to several tracks of existing work reviewed above, our method has three key advantages. {\textbf{Data-adaptivity:}} Our approach is adaptive to arbitrary BM patterns of covariates as well as model misspecification in the control variates, and automatically adjusts to the degree of distributional shifts. 
{{\textbf{Efficiency:}} Our method achieves optimal performance within user-specified classes of control functions broader than those in existing work. Ideally, when the control functions can be fully nonparametrically learned, our proposed estimator achieves the semiparametric efficiency bound.} {\textbf{Computational feasibility and flexibility:}} The control functions can be constructed based on generic machine learning and they are computationally tractable through finite-dimensional quadratic programming, accommodating both parametric and kernel ridge regression. {These grant us a promising trade-off between computational cost and statistical efficiency by seeking to maximize efficiency gains while remaining practically implementable.} 
{\textbf{Incorporation of unlabeled data:}
We develop double/debiased machine learning (DML) operators that effectively exploit abundant unlabeled data while facilitating the use of powerful black-box machine learning algorithms for estimating nuisance functions. We further establish semiparametric efficiency theory for the semi-supervised and BM setting, which justifies the optimality of the proposed DML operators and the resulting estimator under suitable regularity conditions.
}

In addition, we develop a novel screening procedure to {\textbf{identify and exclude misaligned data sources}}. Recognizing that certain distributional shifts between the $\Lsc\Msc_r$ and $\Usc\Csc$ sources can violate the CAM assumption in practical settings, we develop a debiased discrepancy measure method based on the mean squared difference of certain conditional mean model between sources. This screening mechanism prioritizes statistical validity by automatically detecting and excluding BM sources not sufficiently aligned with the target $\mathcal{UC}$ population. Compared to most existing fusion methods that assume perfect transportability, our approach provides protection against bias from misaligned sources while maintaining power to detect truly transportable ones. Also, our approach is rigorously sharper than the commonly used sensitivity analysis strategy based on the mean difference of certain statistics between sources.

\section{Method}\label{sec:method}

\subsection{Preliminary Estimation}\label{sec:method:prelim}
Consider the problem of inferring a linear functional of $\boldsymbol{\gamma}$, denoted by $\c^\top\boldsymbol{\gamma}$, where $\c$ is an arbitrary vector with $\|\c\|_2 = 1$. When the goal is to estimate the full vector $\boldsymbol{\gamma}$, we set $\boldsymbol{c} = \boldsymbol{e}_j$, the $j$-th unit vector in $\mathbb{R}^{p_A}$ where $p_A$ is the dimension of $\A$, i.e., $\boldsymbol{c}^\top \bar{\boldsymbol{\gamma}} = \gamma_j$. We then apply the following method to estimate each $\bar\gamma_j$ for $j = 1, \ldots, p_A$, and combine the resulting scalar estimates to form the final estimator of $\bar{\boldsymbol{\gamma}}$. For predictive applications with a new subject of covariate $\A_{\sf new}\neq \boldsymbol{0}$, one can take $\boldsymbol{c}=\A_{\sf new}/\|\A_{\sf new}\|_2$, estimate $\boldsymbol{c}^\top \bar{\boldsymbol{\gamma}}$, and use $\|\A_{\sf new}\|_2\boldsymbol{c}^\top \bar{\boldsymbol{\gamma}}$ as the linear predictor. Let $\Ehat_{\mathcal{D}}[f(\cdot)]$ denote the empirical mean of a measurable function $f(d)$ over observations $d \in \mathcal{D}$, where $\mathcal{D}\in\{\Lsc\Csc, \Lsc\Msc_1, \dots, \Lsc\Msc_R, \Usc\Csc\}$, e.g., $\Ehat_{\Lsc\Csc}[f] = \frac{1}{n} \sum_{d \in \Lsc\Csc} f(d)$. Similarly, we define $\varhat_{\mathcal{D}}[f]$ as the empirical variance of $f(\cdot)$ on the dataset $\mathcal{D}$, e.g., $\varhat_{\Lsc\Csc}[f] = \frac{1}{n} \sum_{d \in \Lsc\Csc} \left\{ f(d) - \Ehat_{\Lsc\Csc}[f] \right\}^2.$ 
For two sequences $a_n$ and $b_n$, we denote by $a_n=O(b_n)$ if $\lim_{n\rightarrow\infty}|a_n/b_n|<\infty$, $a_n=o(b_n)$ if $\lim_{n\rightarrow\infty}a_n/b_n=0$, and $a_n=\op(b_n)$ or $a_n=\Op(b_n)$ if $a_n=o(b_n)$ or $a_n=O(b_n)$ with a probability approaching $1$.

First, we derive a preliminary estimator for $\bar{\boldsymbol{\gamma}}$ using the $\Lsc\Csc$ and $\Usc\Csc$ samples. Due to the distributional shift of $\X$ between $\Lsc\Csc$ and $\Usc\Csc$, directly solving the standard estimating equation in (\ref{eq:ee:pop}) with $\Lsc\Csc$ will lead to bias. The conventional importance weighting method \citep{gretton2009covariate} adjusts for this bias by reweighting $\Lsc\Csc$ samples with the density ratio $h_{\Lsc\Csc}(\X)$ and can produce a consistent estimator for $\bar{\boldsymbol{\gamma}}$ under assumption (\ref{eq:cs:assume}). However, due to the excessive error of machine learning (ML) estimation for $h_{\Lsc\Csc}(\cdot)$, this strategy tends to fail in achieving $n^{1/2}$-consistency or asymptotic unbiasedness \citep{liu2023augmented}. To address this issue, we introduce the DML operator: {
\begin{equation}
\begin{split}
{\mathfrak{M}}_{\Lsc\Csc}(D)&=h_{\Lsc\Csc}(\X)\big(D-{\rm E}_{\Lsc\Csc}[D\mid\X]\big)+{\rm E}_{\Usc\Csc}\big({{\rm E}}_{\Lsc\Csc}[D\mid\X]\big),
\end{split}
\label{equ:dml}
\end{equation}
where $h_{\Lsc\Csc}(\X)=p_{\Usc\Csc}(\X)/p_{\Lsc\Csc}(\X)$ and $D$ can be any function of observed variables.} Let $\widehat{h}_{\Lsc\Csc}(\X)$ and $\widehat{{\rm E}}_{\Lsc\Csc}[D\mid\X]$ denote ML estimators of the nuisance functions $h_{\Lsc\Csc}(\X)$ and ${{\rm E}}_{\Lsc\Csc}[D\mid\X]$, respectively. We use $\widehat{\mathfrak{M}}_{\Lsc\Csc}(\cdot)$ to represent the empirical version of ${\mathfrak{M}}_{\Lsc\Csc}(\cdot)$, with ${h}_{\Lsc\Csc}$, ${\rm E}_{\Lsc\Csc}[\cdot\mid\X]$, and ${\rm E}_{\Usc\Csc}$ in (\ref{equ:dml}) replaced by their estimators or empirical counterpart $\widehat{h}_{\Lsc\Csc}$, $\widehat{\rm E}_{\Lsc\Csc}[\cdot\mid\X]$, and $\widehat{\rm E}_{\Usc\Csc}$. Then, we obtain the preliminary estimator $\widetilde{\boldsymbol{\gamma}}$ by solving
\begin{equation} \label{eq:glm}
\widetilde{\mathbf{U}}_{\Lsc\Csc}(\boldsymbol{\gamma}) = \Ehat_{\Lsc\Csc}
\Big[\widehat{\mathfrak{M}}_{\Lsc\Csc}\big(\A \{Y - g(\boldsymbol{\gamma}^{\top} \A)\}\big)\Big] = \mathbf{0}.
\end{equation}
The asymptotic expansion of $\widetilde{\boldsymbol{\gamma}}$ is established in Lemma \ref{lem:exp} under the regularity and ML convergence assumptions presented in Appendix. We require the ML estimators of the density ratio and conditional mean models converge to the truth in the rate $\op(n^{-1/4})$.
\begin{lemma}\label{lem:exp}
Under Assumptions A1 -- A2, we have
\[
\sqrt{n}(\widetilde{\boldsymbol{\gamma}} - \bar{\boldsymbol{\gamma}}) =  \sqrt{n}\Ehat_{\Lsc\Csc}{\mathfrak{M}}_{\Lsc\Csc}(\barS) + \op(1),
\]
where $\barS\equiv \barS(\A,Y) = \barJ ^{-1} \A\{Y-g(\bar{\boldsymbol{\gamma}}^\top\A)\}$ and $\barJ =  {\rm E}_{\Usc\Csc}\big[\dot{g}(\bar{\boldsymbol{{\gamma}}}^\top \A) \A \A^\top\big]$. Also, $\sqrt{n}(\widetilde{\boldsymbol{\gamma}} - \bar{\boldsymbol{\gamma}})$ weakly converge to gaussian distribution with mean zero.
\end{lemma}

The preliminary estimation introduced above can be implemented with generic ML methods (e.g., random forests or neural networks) for estimating the nuisance functions. To ensure the asymptotic validity, we employ cross-fitting as used in \citet{chernozhukov2016double}, which avoids overfitting and guarantees asymptotic unbiasedness. To further reduce computational costs, we propose an implementation procedure initializing with the importance weighting estimator using $\widehat{h}_{\mathcal{LC}}(\cdot)$, followed by an one-step Newton-Raphson update based on (\ref{eq:glm}). This one-step estimator can be shown to be asymptotically equivalent to the fully iterative solution of (\ref{eq:glm}), while being computationally more economical. Details of the cross-fitting and one-step estimation are presented in Appendix. For notational simplicity, we drop the subscripts indicating the split data folds in our main paper. One can refer to Appendix for the cross-fitted version of our method used for practical implementation. Our theoretical results also require cross-fitting.

\subsection{Data-adaptive Control Variate Approach}\label{sec:method:cal}

After obtaining the preliminary $\widetilde{\boldsymbol{\gamma}}$, we incorporate the $\Lsc\Msc_r$ samples to reduce its variance. For now, assume the alignment assumption depicted by (\ref{eq:cs:assume}) holds for all sources. The scenario with misaligned sources will be considered in Section \ref{sec:violation}.  {First, we introduce the following operators: 
\begin{equation}
\begin{split}
{\mathfrak{M}}_{r}(D)=&h_{\Lsc\Csc}(\X)\big(D - {{\rm E}}_{\Lsc\Msc_r}[D\mid\Z_r]-{\rm E}_{\Lsc\Csc}[D - {{\rm E}}_{\Lsc\Msc_r}[D\mid\Z_r]\mid\X]\big)\\
&+{\rm E}_{\Usc\Csc}\big({{\rm E}}_{\Lsc\Csc}[D\mid \X]\big),\\
{\mathfrak{M}}_{\Lsc\Msc_r}(D)=&h_{\Lsc\Msc_r}(\Z_r)\big(D-{{\rm E}}_{\Lsc\Msc_r}[D\mid\Z_r]\big) + {\rm E}_{\Usc\Csc}\big({{\rm E}}_{\Lsc\Csc}[{{\rm E}}_{\Lsc\Msc_r}[D\mid\Z_r]\mid \X]\big),    
\end{split}    
\end{equation}
where $h_{\Lsc\Csc}(\X)=p_{\Usc\Csc}(\X)/p_{\Lsc\Csc}(\X)$ and $h_{\Lsc\Msc_r}(\Z_r)=p_{\Usc\Csc}(\Z_r)/p_{\Lsc\Msc_r}(\Z_r)$.} Let $\widehat{h}_{\Lsc\Csc}(\X)$, $\widehat{h}_{\Lsc\Msc_r}(\Z)$,  $\widehat{{\rm E}}_{\Lsc\Csc}[D\mid\X]$ and $\widehat{{\rm E}}_{\Lsc\Msc_r}[D\mid\Z_r]$ denote ML estimators of the nuisance functions $h_{\Lsc\Csc}(\X)$, $h_{\Lsc\Msc_r}(\Z_r)$, ${{\rm E}}_{\Lsc\Csc}[D\mid\X]$ and ${{\rm E}}_{\Lsc\Msc_r}[D\mid\Z_r]$, respectively. Also, let $\widehat{\mathfrak{M}}_{r}$ and $\widehat{\mathfrak{M}}_{\Lsc\Msc_r}$ denote the empirical versions of ${\mathfrak{M}}_{r}$ and ${\mathfrak{M}}_{\Lsc\Msc_r}$. Similar to our proposal in Section \ref{sec:method:prelim}, we use DML on the operators $\widehat{\mathfrak{M}}_r$ and $\widehat{\mathfrak{M}}_{\Lsc\Msc_r}$ to correct bias due to distributional shifts. 

By Lemma \ref{lem:exp}, the asymptotic variance of $\c^\top \widetilde{\boldsymbol{\gamma}}$ is proportional to that of $\c^\top {\mathfrak{M}}_{\Lsc\Csc}(\barS)$. Inspired by the control variate approach \citep{glasserman2004monte}, we augment $\c\trans\widetilde{\boldsymbol{\gamma}}$ as:
\begin{equation}
\widehat{\beta}(\bphi)=\c\trans\widetilde{\boldsymbol{\gamma}}+ \sum_{r=1}^R\left\{\Ehat_{\Lsc\Msc_r}\left[\widehat{\mathfrak{M}}_{\Lsc\Msc_r}(\phi_{r})\right]-\Ehat_{\Lsc\Csc}\left[\widehat{{\mathfrak{M}}}_{r}(\phi_{r})\right]\right\},
\label{equ:semieff}
\end{equation}
where $\phi_{r}\equiv \phi_{r}(\X_{\Gamma_r},Y)$ is some (possibly data-dependent) control function of the observed variables in $\Lsc\Msc_r$ and we denote by $\bphi = (\phi_{1},\dots,\phi_{R})^{\top}$. By the tower rule of conditional expectation, the population version of augmentation terms in (\ref{equ:semieff}) satisfy that for each $r$,
\begin{equation}\label{eq: pop bias}
    {\rm E}_{\Lsc\Msc_r}\left[\mathfrak{M}_{\Lsc\Msc_r}(\phi_{r})\right]-{\rm E}_{\Lsc\Csc}\left[\mathfrak{M}_{r}(\phi_{r})\right]={\rm E}_{\Usc\Csc}\Big({{\rm E}}_{\Lsc\Csc}[{{\rm E}}_{\Lsc\Msc_r}[\phi_r\mid\Z_r]\mid \X]-{{\rm E}}_{\Lsc\Csc}[{{\rm E}}_{\Usc\Csc_r}[\phi_r\mid\Z_r]\mid \X]\Big),
\end{equation}
which equals to zero under the CAM assumption (\ref{eq:cs:assume}). Thus, under the regularity and rate assumptions introduced in Section \ref{sec:theory}, the augmentation term $\Ehat_{\Lsc\Msc_r}[\widehat{\mathfrak{M}}_{\Lsc\Msc_r}(\phi_{r})]-\Ehat_{\Lsc\Csc}[\widehat{\mathfrak{M}}_{r}(\phi_{r})]$ is asymptotically mean-zero and only affect the variance of $\widehat{\beta}(\bphi)$. Consequently, $\widehat{\beta}(\bphi)$ is still valid, while its variance could be potentially smaller than the preliminary estimator $\c\trans\widetilde{\boldsymbol{\gamma}}$ when $\phi_{r}$ is specified properly. {Note that the operators $\mathfrak{M}_{\Lsc\Csc}$, $\mathfrak{M}_{r}$, $\mathfrak{M}_{\Lsc\Msc_r}$ are constructed by incorporating the abundant unlabeled data from $\Usc\Csc$. The semiparametric theory developed in Section~\ref{sec:thm:re} for the semi-supervised setting shows that these operators can efficiently leverage the information contained in the unlabeled data, provided that the control functions are properly specified.}
{\begin{remark}
When $\Z_r$ does not contain $Y$, ${\mathfrak{M}}_{r}$ and ${\mathfrak{M}}_{\Lsc\Msc_r}$
can be simplified as ${\mathfrak{M}}_{r}(D)={\mathfrak{M}}_{\Lsc\Csc}(D)$ and
$
{\mathfrak{M}}_{\Lsc\Msc_r}(D)=h_{\Lsc\Msc_r}(\Z_r)\big(D-{{\rm E}}_{\Lsc\Msc_r}[D\mid\Z_r]\big) + {\rm E}_{\Usc\Csc}\big({{\rm E}}_{\Lsc\Msc_r}[D\mid\Z_r]\big).
$
When $\Z_r=(Y,\X_r)$, i.e., all observed variables on $\Lsc\Msc_r$ are not distributionally aligned with $\Lsc\Csc$, we have ${\mathfrak{M}}_{\Lsc\Msc_r}(D)=0$ for any $D$. That means, $\Lsc\Msc_r$ cannot help with variance reduction, which is natural since no observed variable is aligned between $\Lsc\Msc_r$ and $\Usc\Csc$ in this scenario.
\end{remark}
}

{
Next, we specify the control function $\bphi$ by investigating the efficiency of the estimator $\widehat{\beta}(\bphi)$. For any given function $\bphi$, straightforward asymptotic expansion and derivation can show that the asymptotic variance of $\sqrt{n}(\widehat{\beta}(\bphi) - \c\trans\boldsymbol{\bar{\gamma}})$ is
\begin{equation}\label{eq: variance objective}
\begin{aligned}
    Q(\bphi)
    & = \var_{\Lsc\Csc}\left[{\mathfrak{M}}_{\Lsc\Csc}(\c^\top \barS) -\sum_{r = 1}^{R}{\mathfrak{M}}_{r}(\phi_{r})\right]
    + \sum_{r = 1}^{R}\frac{1}{\rho_{r}}\var_{\Lsc\Msc_r}[{\mathfrak{M}}_{\Lsc\Msc_r}(\phi_{r})].
\end{aligned}
\end{equation}

Motivated by this, we propose to obtain $\bphi$ by minimizing the empirical version of $Q(\bphi)$ with respect to $\bphi$ over a function class $\cF_{n}$ that may depend on the sample size $n$. 
By doing so, our estimator can achieve {\em intrinsic efficiency} \citeg{tan2010bounded,rotnitzky2012improved} over the control variate estimators with a certain class of control function. Moreover, the semiparametric theory developed in Section~\ref{sec:thm:re} suggests that the estimator $\widehat{\beta}(\bphi)$ attains {\em semiparametric efficiency} under suitable regularity conditions provided that the function class $\cF_{n}$ is sufficiently rich. 
Let $\widetilde{\S}\equiv\widetilde{\S}(\A,Y)$ be an estimate of $\barS(\A,Y)$ with $\bar{\boldsymbol{\gamma}}$ replaced by the preliminary $\widetilde{\boldsymbol{\gamma}}$. 
We then approximate $Q(\bphi)$ by
$
\widehat{Q}(\bphi)=\varhat_{\Lsc\Csc}\left[\widehat{\mathfrak{M}}_{\Lsc\Csc}\Big(\c^\top\widetilde{\S}\Big)-\sum^R_{r=1}\widehat{\mathfrak{M}}_{r}(\phi_r)\right]+\sum^R_{r=1}\rho_r^{-1}\varhat_{\Lsc\Msc_r}\left[\widehat{\mathfrak{M}}_{\Lsc\Msc_r}(\phi_r)\right]$.
The control function is then obtained via the penalized minimization problem:
\begin{equation}
\tbphi=\arg\min_{\bphi\in \cF_{n}}\left\{
\widehat{Q}(\bphi) + \zeta_{\blambda}(\bphi)
\right\},
\label{eq: emp obj}
\end{equation}
where $\zeta_{\blambda}$ is some regularization term that may depend on the function class $\cF_n$ and $\blambda$ is a vector of tuning parameters. Then, the DEFUSE estimator is constructed as
\begin{equation}
\widehat{\beta}_{\DEF} = \widehat{\beta}(\tbphi) = \c\trans\widetilde{\boldsymbol{\gamma}}+ \sum_{r=1}^R\left\{\Ehat_{\Lsc\Msc_r}\left[\widehat{\mathfrak{M}}_{\Lsc\Msc_r}(\widetilde{\phi}_{r})\right]-\Ehat_{\Lsc\Csc}\left[\widehat{{\mathfrak{M}}}_{r}(\widetilde{\phi}_{r})\right]\right\}.
\label{eq: defuse def}
\end{equation}

Note that $Q(\bphi)$ involves multiple conditional mean evaluations. Those conditional mean can be estimated either by (i) directly regressing the functions $\c^\top\widetilde{\S}$ or $\phi_r$ against $\X$ or $\{\X_{\Gamma_r},Y\}$ weighted by $h_{\Lsc\Csc}(\X)$ on $\Lsc\Csc$, or (ii) learning the conditional distributions $Y\mid \X$, $\X_{\Gamma_r^c}\mid (\X_{\Gamma_r}, Y)$, and computing the conditional means via Monte Carlo methods. Crucially, the unbiasedness of the resulting control variate estimator is preserved even if the control function does not converge to the nonparametric global minimizer of $Q(\bphi)$. Unlike typical regression tasks targeting conditional means, \eqref{eq: emp obj} may not be computationally feasible or, at least, sophisticated to solve in practice when $\bphi$ is taken as a generic ML model. This is due to the presence of multiple conditional mean calculations in the operators as well as the addition of different $\phi_{r}$'s supported on different sets of variables in \eqref{eq:  emp obj}. To address this,
we provide a strategy to construct the candidate function class $\cF_{n}$, which leads to quadratic $\widehat{Q}(\bphi)$ and hence facilitates the computation.

Our strategy is to construct the control function $\tbphi$ by refining an ``anchor'' control function $\widehat{\bphi}$, which is either prespecified or obtained by training a black-box algorithm on an independent sample. When $\X$ and $Y$ are observed in all sources, a natural choice of control function is $\c^{\top}\barS(\A,Y)$. Under our setting, however, only $(\X_{\Gamma_r},Y)$ is observed in $\Lsc\Msc_r$, so $\c^{\top}\barS(\A,Y)$ is generally unavailable and one may use an AI model to predict $\c^{\top}\barS(\A,Y)$ from $(\X_{\Gamma_r},Y)$ within $\Lsc\Msc_r$. Such predictors may involve high-dimensional auxiliary features; see simulation setting (VI) in Section~\ref{sec:simu} and the real-data example in Section~\ref{sec:real:mimic}. 

Denote the resulting predictor of $\c^{\top}\barS(\A,Y)$ based on $(\X_{\Gamma_r},Y)$ by $\widehat{\phi}_r$, and let $\widehat{\bphi}=(\widehat{\phi}_1,\dots,\widehat{\phi}_R)^\top$.
Directly using $\widehat{\bphi}$ as the control function may lead to unsatisfactory performance for two reasons: first, the AI predictors may be inaccurate; second, even if they are accurate, they are not designed to minimize the variance criterion $Q(\bphi)$. To overcome this difficulty, we develop a data-adaptive refinement procedure. Let $\W_r\subseteq\{\X_{\Gamma_r},Y\}$ be a prespecified vector of variables. The main idea is to reweight the anchor predictor $\widehat{\phi}_r$ through a function $w_r(\W_r;\bdelta_r)$, and define the function class in \eqref{eq: emp obj} as
$\cF_n=\big\{\bphi(\X,Y; \bdelta) =(w_r(\W_r; \boldsymbol{\delta}_r)\widehat{\phi}_r(\X_{\Gamma_r}, Y))_{r=1,\ldots,R}: \bdelta \in \Csc_{n}\big\}$. Here, $w_r(\cdot;\bdelta_r)$ is a prespecified model linear in $\bdelta_r$, and $\bdelta=(\bdelta_1^\top,\dots,\bdelta_R^\top)^\top$ collects all model parameters constrained to lie in $\Csc_n$. In practice, when the anchor control function $\widehat{\bphi}$ is not readily available, one may simply take $\widehat{\bphi}$ to be a vector of ones. In that case, the proposed method reduces to minimizing the penalized asymptotic variance over a prespecified linear function class.}

We recommend in Examples \ref{example:const}--\ref{example:krr} specific choices for $w_{r}(\cdot)$, covering both parametric and nonparametric models. Notably, all of them can be directly computed via finite-dimensional quadratic programming. Example \ref{example:const} simply uses constant coefficients to re-allocate the data sources $\Lsc\Msc_r$'s, which has a similar spirit as some recent advances \citeg{angelopoulos2023ppi++,miao2023assumption}. Compared to this, Examples \ref{example:line} and \ref{example:krr} enable more powerful adjustment as demonstrated in Section \ref{sec:simu}. Example \ref{example:line} specifies $w_{r}(\W_r)$ as a linear function and Example \ref{example:krr} makes nonparametric optimization computationally tractable via kernel ridge regression (KRR). 

\begin{example}[Simple allocation]
\label{example:const}
Set $w_{r}(\W_r; \delta_r)=\delta_r$ with $\delta_r\in\mathbb{R}$. No penalty is needed in \eqref{eq: emp obj} in this case.
\end{example}

\begin{example}[Linear function]
\label{example:line}
Set $w_{r}(\W_r; \boldsymbol{\delta}_r)=(1,\W_r\trans)\bdelta_r$ with $\bdelta_r\in\mathbb{R}^{{\rm dim}(\W_{r}) +1}$. No penalty is needed in \eqref{eq: emp obj} in this case.
\end{example}

\begin{example}[Kernel ridge regression]
\label{example:krr}
Set $w_{r}(\W_r; \boldsymbol{\delta}_r)$ to be from a RKHS induced by a symmetric and positive semi-definite kernel $K_r(\cdot,\cdot)$ on the domain of $\W_r$. In implementation, $w_{r}(\w; \boldsymbol{\delta}_r)$ can be specified as $\delta_{r,0} + n^{-1}\sum_{i=1}^n \delta_{r,i} K_r(\w,\W_{r,i})
$,
where $\W_{r,i}$ represents the $i$-th observation of $\W_r$ in $\Lsc\Csc$. Here we introduce the intercept term $\delta_{r,0}$ to ensure that $w_{r}(\w; \boldsymbol{\delta}_r) = 1$ for some $\boldsymbol{\delta}_r$. The penalty in \eqref{eq: emp obj} can be set as $\sum_{r = 1}^{R}\lambda_r\|n^{-1}\sum_{i=1}^n \delta_{r,i} K_r(\cdot,\W_{r,i})\|_{K_r}^2$ where $\|\cdot\|_{K_r}$ is the norm in the RKHS space induced by $K_r$.
\end{example}

\subsection{Exclude Misaligned Data Sources}\label{sec:violation}

{
Prior to integrating the labeled data sources, it is crucial to assess their transferability to the target population $\mathcal{UC}$, i.e., whether the CAM assumption in (\ref{eq:cs:assume}) holds. To construct a valid screening procedure without circularity, we set the sample $\mathcal{LC}$ as an internal benchmark. In practice, $\mathcal{LC}$ and $\mathcal{UC}$ are typically collected concurrently under a unified protocol with fully-observed covariates. Thus, it is reasonable to assume that the conditional alignment of $Y\mid \X$ holds securely between these two sources. We leverage this secure alignment to estimate the necessary conditional distributions without relying on the alignment properties of the $\mathcal{LM}_r$ sources. Different from $\mathcal{LC}$, the BM sources $\mathcal{LM}_r$'s may originate from historical studies, external databases, or different institutions where specific features were not collected or defined consistently, leading to the BM pattern. Thus, we leverage $\mathcal{LC}$ as an ``anchor'' source to identify $\mathcal{LM}_r$'s exhibiting significant deviation from the CAM assumption and thus should be excluded from data fusion to avoid bias. 
}

In (\ref{equ:semieff}), the influence from $\Lsc\Msc_r$ is the control variate $\widetilde{V}_r:=\Ehat_{\Lsc\Msc_r}\big[\widehat{\mathfrak{M}}_{\Lsc\Msc_r}\big(\phi_r\big)\big]-\Ehat_{\Lsc\Csc}\big[\widehat{\mathfrak{M}}_{r}\big(\phi_r\big)\big]$, which converges to $\bar{V}_r:={\rm E}_{\Lsc\Msc_r}\left[\mathfrak{M}_{\Lsc\Msc_r}(\phi_{r})\right]-{\rm E}_{\Lsc\Csc}\left[\mathfrak{M}_{r}(\phi_{r})\right]$. Motivated by our above discussion, we assume the covariate shift assumption for $\Lsc\Csc$ in (\ref{eq:cs:assume}) safely holds. Then by \eqref{eq: pop bias}, $\bar{V}_r = \mathrm{E}_{\Usc\Csc}\big[m_r(\X)- m^*_r(\X)\big]$
where 
$m_r(\X)={{\rm E}}_{\Lsc\Csc}[{{\rm E}}_{\Lsc\Msc_r}[\phi_r\mid\Z_r]\mid \X]$ and \(m_r^*(\X)={\rm E}_{\Lsc\Csc}\left[{\rm E}_{\Lsc\Csc}[h_{\Lsc\Csc}(\X)\phi_r\mid\Z_r]/{{\rm E}_{\Lsc\Csc}[h_{\Lsc\Csc}(\X)\mid\Z_r]}\mid\X\right]\). When the CAM assumption in  (\ref{eq:cs:assume}) fails for $\Lsc\Msc_r$, it is likely that $\bar{V}_r\neq 0$, which induces bias to our estimator. In this situation, it is wise to exclude $\Lsc\Msc_r$. 

To effectively identify such bias-prone sources, we introduce the mean squared discrepancy (MSD) measure:
$
\bar{\Delta}_r^2=\mathrm{E}_{\Usc\Csc}\big[m_r(\X)- m_r^*(\X)\big]^2
$.
We have $\bar{\Delta}_r=0$ when the CAM assumption (\ref{eq:cs:assume}) holds for $\Lsc\Msc_r$; otherwise, $\bar{\Delta}_r$ tends to be larger than zero. Thus, we aim to exclude $\Lsc\Msc_r$ from data fusion if $r\in\mathcal{A}^c$ where the set of aligned sources $\mathcal{A}=\{r:\bar{\Delta}_r=0\}$ and $\mathcal{A}^c=\{1,\ldots,R\}\setminus\mathcal{A}$. Importantly, since $|\bar{V}_r|\leq \bar{\Delta}_r$, $\bar{\Delta}_r$ is a sharper and more sensitive measure for the violation of CAM compared to the mean difference $\bar{V}_r$ commonly used for sensitivity analysis in data fusion \citeg{cheng2021adaptive,han2025federated}. 

{For any data function $D$, Let $\widehat\Pi_{\Usc\Csc}(D\mid \Z_r) = \widehat{\rm E}_{\Lsc\Csc}[\widehat h_{\Lsc\Csc}(\X)D\mid\Z_r]/\widehat{\rm E}_{\Lsc\Csc}[\widehat h_{\Lsc\Csc}(\X)\mid\Z_r]$  be an ML estimate of ${\rm E}_{\Usc\Csc}[D\mid \Z_r]$.
Empirically, we estimate $m_r(\X)$ and $m^*_r(\X)$ by  $\widehat{m}_r(\X) = \Ehat_{\Lsc\Csc}[\Ehat_{\Lsc\Msc_r}[\phi_r\mid\Z_r]\mid \X]$ and $\widehat m_r^*(\X)=\widehat{\rm E}_{\Lsc\Csc}\big[\widehat\Pi_{\Usc\Csc}(\phi_r\mid \Z_r)\mid\X\big]$. Simply plugging in $\widehat{m}_r(\cdot)$ and $\widehat{m}^*_r(\cdot)$ to estimate MSD could result in excessive bias that dominates the actual $\bar{\Delta}_r$. To address this issue, we propose a debiasing method. Suppose $\mathcal{G}$ is a prespecified function space satisfying (i) $\mathrm{E}_{\Usc\Csc}[d_r^2(\X)]\leq 1$ for all $d_r\in\mathcal{G}$ and (ii) $d_r^*\in\mathcal{G}$, where $d_r^*(\X):=\{m_r(\X)- m_r^*(\X)\}/\{\mathrm{E}_{\Usc\Csc}[m_r(\X)- m_r^*(\X)]^2\}^{1/2}$ if $m_r(\X)- m_r^*(\X)\neq 0$ and $d_r^*(\X):=1$, otherwise. Then we have
\begin{equation}
\bar{\Delta}_r=\max_{d\in\mathcal{G}}L_r(d_r),\quad \mbox{where}\quad L_r(d
_r)=\mathrm{E}_{\Usc\Csc}\left[d_r(\X)\big\{m_r(\X)- m_r^*(\X)\big\}\right].
\label{equ:def:delta}
\end{equation}

We construct the debiased estimator for the objective function $L_r(d_r)$:
\begin{equation}
\begin{split}
\widehat L_r(d_r)
&=\Ehat_{\Usc\Csc}\left[d_r(\X)\{\widehat m_r(\X)-\widehat m_r^*(\X)\}\right]\\
&+\Ehat_{\Lsc\Msc_r}\left[\widehat h_{\Lsc\Msc_r}(\Z_r)\widehat\Pi_{\Usc\Csc}(d_r\mid \Z_r) \left\{\phi_r-\widehat{\rm E}_{\Lsc\Msc_r}[\phi_r\mid\Z_r]\right\}\right]\\
&+\Ehat_{\Lsc\Csc}\left[\widehat h_{\Lsc\Csc}(\X)d_r(\X)\left\{\widehat{\rm E}_{\Lsc\Msc_r}[\phi_r\mid\Z_r]-\widehat m_r(\X)\right\}\right]\\
&-\Ehat_{\Lsc\Csc}\left[\widehat{\mathfrak{M}}_{\Lsc\Csc}\left(\widehat\Pi_{\Usc\Csc}(d_r\mid \Z_r)\left\{\phi_r-\widehat\Pi_{\Usc\Csc}(\phi_r\mid \Z_r)\right\}\right)\right]\\
&-\Ehat_{\Lsc\Csc}\left[\widehat h_{\Lsc\Csc}(\X)d_r(\X)\left\{\widehat\Pi_{\Usc\Csc}(\phi_r\mid \Z_r)-\widehat m_r^*(\X)\right\}\right].
\end{split}
\label{equ:l:debias}
\end{equation}}
Based on \eqref{equ:l:debias}, we estimate \(\bar\Delta_r\) through
\begin{equation}
\widehat{\Delta}_r=\widehat{L}_r(\widehat{d_r}),\quad\mbox{where} \quad \widehat{d}_r=\arg\max_{d_r\in\mathcal{G}}\widehat{L}_r(d_r).
\label{equ:delta:emp}
\end{equation}
This enables consistent identification of misaligned $\Lsc\Msc_r$ with $\bar{\Delta}_r\gg \{\xi(n)/n\}^{1/2}$, where $\xi(n)$ is a model complexity parameter of the model space $\mathcal{G}$ that may depend on $n$; see Section \ref{sec:thm:screen} for details. Based on this, we take the well-aligned set of the BM sources as $\widehat{\Acal}=\{r:\widehat{\Delta}_r<\tau\}$, where the threshold parameter $\tau = \{\xi(n)/n\}^{1/2-\epsilon}$ for some small $\epsilon>0$, saying $\epsilon=0.02$ in practice. Only those $\Lsc\Msc_r$ sources with $r\in\widehat{\Acal}$ will be used for the data fusion in Section \ref{sec:method:cal}. Theoretically, we should consider arbitrary measurable function to verify CAM condition but in practice we consider the control functions $\phi_r$'s for implementation. We first derive $\phi_r$'s using all BM sources then base on them to screen and exclude the misaligned BM sources, and repeat this two steps until no BM source is diagnosed as misaligned anymore. Again, to avoid overfitting bias, we require cross-fitting of all nuisance ML models, including the DML operators and the control functions.

\section{Asymptotic Analysis}\label{sec:theory}

\subsection{Asymptotic properties of DEFUSE}

In this section, we establish the asymptotic properties of DEFUSE. We assume that $\rho_r=n_r/n$ converges to some $\bar\rho_r\in(0,\infty)$, $n+ \sum^R_{r = 1}n_r=o(N)$. The main assumptions are introduced in Appendix. In summary, Assumption A1 includes mild regularity conditions on the distribution of the variables, overlapping (positivity) assumption between data sources, and positive definiteness of the hessian matrix. Assumption A2 imposes that the nuisance ML estimators $\widehat{h}_{\Lsc\Csc}(\cdot)$, $\widehat{h}_{\Lsc\Msc_r}(\cdot)$,  $\widehat{{\rm E}}_{\Lsc\Csc}[\cdot\mid\X]$ and $\widehat{{\rm E}}_{\Lsc\Msc_r}[\cdot\mid\Z_r]$ converge to the corresponding true functions with their $L_2$-errors being $\op(n^{-1/4})$, which is known as the DML or rate doubly robust assumption \citep{chernozhukov2016double}. Assumption A3 imposes the consistency of the empirical control functions in $\widetilde{\bphi}$. Importantly, we only require their $\op(1)$-convergence to limiting functions denoted as $\bar{\bphi}$. 

For $\widetilde{\bphi}$ constructed following the RKHS Example \ref{example:krr}, we justify its consistency in Proposition \ref{prop:phi_tilde_convergence_joint} under standard assumptions for the analysis of KRR. Also, both Examples \ref{example:const} and \ref{example:line} are more simple and can be viewed as special cases of Example \ref{example:krr} with constant and linear kernels. Thus, Proposition \ref{prop:phi_tilde_convergence_joint} provides a primitive sufficient condition for Assumption A3 under our recommended construction strategies for the control functions. For any probability measure $Q$, we define the $L_2(Q)$ norm of a possibly random measurable scalar-valued function $h$ as
$\|h\|_{L_2(Q)} = \Big\{\int h^2(z) dQ(z)\Big\}^{1/2}$. Define the limit of $\mathcal{F}_{n}$ as 
$\mathcal{F} = \big\{\boldsymbol{f} \in \mathcal{H}: \lim_{n\to\infty}\|\boldsymbol{f}_{n} - \boldsymbol{f}\|_{\mathcal{H}} = 0\ \text{in probability for some sequence $\{\boldsymbol{f}_{n}\}_{n = 1}^{\infty}$ such that $\boldsymbol{f}_{n} \in \mathcal{F}_{n}$}\big\}$.
\begin{prop}
\label{prop:phi_tilde_convergence_joint}
Under Assumption~A8 in Appendix, the control functions $\widetilde\bphi$ constructed under Example \ref{example:krr} satisfy that $\sum_{r=1}^R
\|\widetilde\phi_r-\bar{\phi}_{r}\|_{L_{2}(P_{\Lsc\Msc_r})}
=
o_p(1)$ for some deterministic functions $\bar{\phi}_r$ ($r = 1,\dots, R$) such that $Q(\bar{\bphi}) = \inf_{\bphi \in \cF}Q(\bphi)$, where $\bar{\bphi} = (\bar{\phi}_1, \dots, \bar{\phi}_R)^{\top}$.
\end{prop}

\begin{remark}
Notably, our required convergence rates on the nuisance functions differ substantially between Assumptions A2 and A3. For the density ratio and conditional mean models with the input $\X$ or $\Z_r$, $\op(n^{-1/4})$-convergence to the truth is required such that the DML constructions $\widehat{\mathfrak{M}}_{\Lsc\Csc}(\cdot)$, $\widehat{\mathfrak{M}}_{\Lsc\Msc_r}(\cdot)$ can adjust for distributional shifts and ensure $n^{1/2}$-consistency. In contrast, for the control functions, we only require a much less stringent $\op(1)$ rate, thanks to our DML construction adjusting for distributional shifts of the control variates. This is a blessing for the incorporation of high-dimensional or AI-derived features into the control functions as mentioned in Remark \ref{rem:intro}. 
\end{remark}


Theorem \ref{thm:1} establishes that the DEFUSE estimator $\widehat{\beta}_{\DEF}$ is $\sqrt{n}$-consistent, asymptotically unbiased, and asymptotically normal under Assumptions~A1--A3. Based on this, the efficiency properties of $\widehat{\beta}_{\DEF}$ are established next. Throughout the theoretical development, we assume $\mathcal{F}_{n}$ is a possibly random subspace of a functional space $\mathcal{H}$ with norm $\|\cdot\|_{\mathcal{H}}$. 

\begin{theorem}[Convergence of DEFUSE]
Under Assumptions A1--A3 and when the CAM conditions are satisfied, $\widehat{\beta}_{\DEF} \overset{p}{\to} \c\trans\boldsymbol{\bar{\gamma}}$ and 
\begin{align*}
\sqrt{n}(\widehat{\beta}_{\DEF}-\c^\top\bar{\boldsymbol{\gamma}})
&=\sqrt{n}\Ehat_{\Lsc\Csc}{\mathfrak{M}}_{\Lsc\Csc}(\c^\top\barS)+\sqrt{n}\sum_{r=1}^R\left\{\Ehat_{\Lsc\Msc_r}\left[{\mathfrak{M}}_{\Lsc\Msc_r}(\bar{\phi}_{r})\right]-\Ehat_{\Lsc\Csc}\left[{{\mathfrak{M}}}_{r}(\bar{\phi}_{r})\right]\right\}+ \op(1),
\end{align*}
where $\barS$ is as defined in Lemma \ref{lem:exp} and $\bar\bphi=(\bar{\phi}_1, \dots, \bar{\phi}_R)^{\top}$ is the vector of deterministic functions defined in Assumption A3. Consequently, $\sqrt{n}(\widehat{\beta}_{\DEF}-\c^\top\bar{\boldsymbol{\gamma}})$ weakly converges mean zero normal distribution with asymptotic variance $Q(\bar\bphi)$
.
\label{thm:1}
\end{theorem}

\subsection{Efficiency Properties}\label{sec:thm:re}

{In this section, we study the semiparametric theory for the setup considered in this paper to investigate the optimality property of our proposed estimator. Suppose that the distribution of $\X$ can be estimated with high precision since the sample size from $\Usc\Csc$ is much larger than that of the labeled sources in our semi-supervised (SS) framework. To characterize this setting, we consider the semi-supervised semiparametric framework similar to \citet{cheng2021robust}, in which $p_{\Usc\Csc}(\X)$ is assumed to be known and uncertainty arising from the estimation on $\Usc\Csc$ can be ignored. Extending our results to the case where $p_{\Usc\Csc}(\X)$ is unknown can be achieved by accounting for the additional randomness introduced by the unlabeled data. With a slight abuse of notation, let the source indicator $\mathcal{D} \in \{\Lsc\Csc, \Lsc\Msc_{1},\dots,\Lsc\Msc_{R}\}$ be a random variable that indicates which source an observation in $\{\Lsc\Csc, \Lsc\Msc_{1},\dots,\Lsc\Msc_{R}\}$ actually comes from, and use $\rho_r$ to denote $\Pr(\mathcal{D} = \Lsc\Msc_r)/\Pr(\mathcal{D} = \Lsc\Csc)$. We establish in Theorem \ref{prop: EIF} the semiparametric efficiency of our estimator defined in (\ref{eq: defuse def}).
\begin{theorem}[Semiparametric efficiency]
\label{prop: EIF}
    Under the CAM assumption in equation \eqref{eq:cs:assume}, $\boldsymbol{c}^\top \bar{\boldsymbol{\gamma}}$ is pathwise differentiable and the influence functions of $\boldsymbol{c}^\top \bar{\boldsymbol{\gamma}}$ can be formed as
    \[
    \begin{aligned}
        {\rm IF}(\bphi) = &\frac{1\{\mathcal{D} = \Lsc\Csc\}}{\Pr(\mathcal{D} = \Lsc\Csc)}\left\{{\mathfrak{M}}_{\Lsc\Csc}(\c^\top \barS) - \sum_{r = 1}^{R}{\mathfrak{M}}_{r}(\phi_{r})\right\} +
        \sum_{r = 1}^{R}\frac{1\{\mathcal{D} = \Lsc\Msc_r\}}{\Pr(\mathcal{D} = \Lsc\Msc_r)}{\mathfrak{M}}_{\Lsc\Msc_r}(\phi_{r}),
    \end{aligned}
    \]
    where $\bphi = (\phi_{1}, \dots, \phi_{R})^{\top}$ and $\phi_{r}$ is some function of $(\X_{\Gamma_r}, Y)$ for $r = 1,\dots, R$ that is mean zero and square integrable under the distribution of $\Usc\Csc$. The influence function is the efficient one if $\phi_{r}(\X_{\Gamma_r}, Y)$ is set as   
    \[
    \phi_{r}^{*}(\X_{\Gamma_r}, Y) = \frac{\Pr(\mathcal{D} = \Lsc\Msc_r)}{h_{\Lsc\Msc_r}(\Z_r)}{\rm E}_{\Usc\Csc}(b\mid \X_{\Gamma_r}, Y) - {\rm E}_{\Usc\Csc}\left[\frac{\Pr(\mathcal{D} = \Lsc\Msc_r)}{h_{\Lsc\Msc_r}(\Z_r)}{\rm E}_{\Usc\Csc}(b\mid \X_{\Gamma_r}, Y)\right],
    \]
    where $b$ is a solution of the integral equation
    \begin{equation}\label{eq: integral equation}
        \begin{aligned}
        &\frac{\Pr(\mathcal{D} = \Lsc\Csc)}{h_{\Lsc\Csc}(\X)}\{b - {\rm E}_{\Lsc\Csc}(b\mid \X)\} + \sum_{r = 1}^{R}\frac{\Pr(\mathcal{D} = \Lsc\Msc_r)}{h_{\Lsc\Msc_r}(\Z_r)}\big\{{\rm E}_{\Usc\Csc}(b\mid \X_{\Gamma_r}, Y) - {\rm E}_{\Usc\Csc}(b\mid \Z_r)\big\}\\
        & - \sum_{r = 1}^{R}{\rm E}_{\Lsc\Csc}\left[\frac{\Pr(\mathcal{D} = \Lsc\Msc_r)}{h_{\Lsc\Msc_r}(\Z_r)}\big\{{\rm E}_{\Usc\Csc}(b\mid \X_{\Gamma_r}, Y) - {\rm E}_{\Usc\Csc}(b\mid \Z_r)\big\}\mid \X\}\right] = \c^\top \barS - {\rm E}_{\Lsc\Csc}(\c^\top \barS\mid \X).
    \end{aligned}
    \end{equation}
    Moreover, $\bphi^{*} = (\phi_{1}^{*},\dots,\phi_{R}^{*})^{\top}$ minimizes the variance
    \[
    \var\{{\rm IF}(\bphi)\} = \frac{Q(\bphi)}{\Pr(\mathcal{D} = \Lsc\Csc)} = \frac{\var_{\Lsc\Csc}\left[{\mathfrak{M}}_{\Lsc\Csc}(\c^\top \barS) -\sum_{r = 1}^{R}{\mathfrak{M}}_{r}(\phi_{r})\right]}{\Pr(\mathcal{D} = \Lsc\Csc)}
    + \sum_{r = 1}^{R}\frac{\var_{\Lsc\Msc_r}[{\mathfrak{M}}_{\Lsc\Msc_r}(\phi_{r})]}{\Pr(\mathcal{D} = \Lsc\Msc_r)}
    \]
    over the set of all square integrable functions.
\end{theorem}
Note that ${\rm IF}(\bphi)$ in
Theorem \ref{prop: EIF} corresponds to the form of the control variate estimator $\widehat{\beta}(\bphi)$ defined in (\ref{equ:semieff}). Theorem \ref{prop: EIF} implies that any regular asymptotically linear estimator for $\boldsymbol{c}^\top \bar{\boldsymbol{\gamma}}$ is asymptotically equivalent to some $\widehat{\beta}(\bphi)$, and the semiparametric efficiency bound can be achieved if the control function $\bphi$ is taken as $\bphi^{*}$ that nonparametrically minimizes $Q(\bphi)$. Thus, our estimator $\widehat{\beta}_{\DEF}$ is semiparametric efficient when the limiting function space $\mathcal{F}$ defined above Proposition \ref{prop:phi_tilde_convergence_joint} is broad enough to include the optimal $\bphi^{*}$.

}

{Due to the limited sample sizes in practice, the control function class $\mathcal{F}_n$ used in (\ref{eq: emp obj}) and its limit $\cF$ are often specified without sufficiently high structural complexity to include the optimal $\bphi^{*}$ achieving semiparametric efficiency. For example, our recommended Examples \ref{example:const} and \ref{example:line} restrict to linear parametric functions and in Example \ref{example:krr}, $\bphi^{*}$ may belong to some less-smooth interpolation space than the user-specified RKHS. Moreover, when $\X_{\Gamma_r}$ is high-dimensional as discussed in Remark \ref{rem:intro}, most covariates can only be included in the control function through some AI-derived predictor $\widehat{\bphi}$ and our reweighting function $w_{r}(\W_r; \boldsymbol{\delta}_r)$ only includes a small number of key features in $\W_r$ but not the entire $(\X_{\Gamma_r},Y)$.} In this situation, we justify the intrinsic efficiency of DEFUSE and study its relative efficiency (RE) compared to the preliminary estimators. Define $\mbox{RE}(\widehat{b}_1|\widehat{b}_2) = {\rm AVar}(\widehat{b}_2)/{\rm AVar}(\widehat{b}_1)$ as the RE between two asymptotically normal estimators $\widehat{b}_1$ and $\widehat{b}_2$, where ${\rm AVar}(\cdot)$ represents their asymptotic variances.

\begin{coro}[Intrinsic efficiency and adaptivity]
\label{coro:1}
Under Assumptions A1--A3 and deterministic limit $\bar\bphi\in\mathop{\arg\min}_{\bphi\in\mathcal{F}}Q(\bphi)$, $\widehat{\beta}_{\DEF} = \widehat{\beta}(\tbphi)$ defined in (\ref{eq: defuse def}) is intrinsically efficient, i.e., having the smallest asymptotic variance being $\min_{\bphi\in\mathcal{F}}Q(\bphi)$, among the class of estimators $\{\widehat{\beta}(\bphi):\bphi\in\mathcal{F}\}$, where $Q(\bphi)$ is defined in (\ref{eq: variance objective}). Consequently, when Assumption A4 further holds (constants $1$ and $0$ belong to feasible sets of $w_r(\cdot)$), $\mbox{RE}\big(\widehat{\beta}_{\DEF}|\c\trans\widetilde{\boldsymbol{\gamma}}\big)\geq 1$ and $ \mbox{RE}\big(\widehat{\beta}_{\DEF}|\widehat{\beta}(\bphi^{\dagger})\big)\geq 1$, where $\bphi^{\dagger}$ is the probability limit of the anchor control function $\widehat{\bphi}$.
\end{coro}


\subsection{Screening Misaligned Sources}\label{sec:thm:screen}

{
In this section, we study the theoretical property of the screening method introduced in Section \ref{sec:violation}. We introduce two additional assumptions. Assumption~A5 imposes that the complexity of the function class $\mathcal{G}$ used in (\ref{equ:def:delta}) is controlled by some effective dimensionality parameter $\xi(n)$, which is commonly used for studying the empirical process \citeg{van1996weak}. Assumption~A6 requires the misaligned sources indexed by $\mathcal{A}^c$ exhibit sufficiently large discrepancy from the $\Usc\Csc$ target benchmark identified through $\Lsc\Csc-\Usc\Csc$ alignment, specifically \(
\inf_{r\in\mathcal A^c}
\bar\Delta_r
\ge
C\left\{\frac{\xi(n)}{n}\right\}^{1/2-c}.
\) with some $c>0$, for all $r\in\mathcal{A}^c$. This ensures that the sources aligned with respect to the MSD criterion ($\bar{\Delta}_r=0$) and the misaligned ones ($\bar{\Delta}_r>0$) can be well-separated by a proper threshold on $\bar{\Delta}_r$. Theorem \ref{thm:3} shows that our screening method can correctly exclude them while preserving the aligned ones, with probability approaching one. 

\begin{theorem}[Selection consistency for aligned sources]\label{thm:3}
Under Assumptions A1--A3, A5, A6 and assuming $d_r^*(\X):=\{m_r(\X)- m_r^*(\X)\}/\{\mathrm{E}_{\Usc\Csc}[m_r(\X)- m_r^*(\X)]^2\}^{1/2}\in\mathcal{G}$ for all $r\in\mathcal{A}^c$, there exists $\epsilon\in(0,c)$ such that screening set
$
\widehat{\mathcal{A}} = \big\{r:\widehat{\Delta}_r <\{\xi(n)/n\}^{1/2-\epsilon} \big\}
$
is consistent with the true $\mathcal{A}:=\{r:\bar{\Delta}_r=0\}$, i.e., $\Pr(\widehat{\mathcal{A}}=\mathcal{A})\to 1$ as $n\to\infty$. 
\end{theorem}
}

\begin{remark}
\label{rem:4}
Theorem \ref{thm:3} considers a simplified setup directly assuming the truth $d_r^*\in \mathcal{G}$. For methods like kernel regression, $\mathcal{G}$ used in (\ref{equ:delta:emp}) does not exactly contain $d_r^*(\cdot)$ but approximates it as $n\rightarrow\infty$. Our analyses can be generalized to such scenarios. When the sample sizes are relatively small, one may choose $\mathcal{G}$ in (\ref{equ:delta:emp}) with low complexity (e.g., low-dimensional parametric models) for the sake of bias-variance tradeoff, even though it may be misspecified. Meanwhile, the commonly used mean-difference strategy \citeg{cheng2021adaptive,miao2023assumption,han2025federated} actually corresponds to setting the function space $\mathcal{G}=\{1\}$ in our framework, which may be too simple to capture the misalignment compared to a more properly specified $\mathcal{G}$ in our method.
\end{remark}

The consistency in Theorem \ref{thm:3} may not hold when there exist misaligned sources with $\bar{\Delta}_r\asymp \{\xi(n)/n\}^{1/2}$ (contradicting Assumption~A6). In the presence of misaligned sources with insufficient discrepancy, our framework cannot ensure valid and data-fused asymptotically normal inference. A similar impossibility result was established in \citet{chen2025minimax} on the fusion of experimental and observational studies. 
One may incorporate non-normal inference approaches \citep{yang2019combining,guo2025robust} to mitigate this.

\section{Simulation Study}\label{sec:simu}

We conduct simulation studies to assess the performance of DEFUSE, in comparison with a broad range of existing methods. Table~\ref{tab:simu:setting} summarizes all simulation settings including various data generation and missing mechanisms. Specifically, we distinguish between the MCAR and CAM mechanisms and consider both the binary and continuous response scenarios. 
We also include the single BM ($R=1$) and multiple BM ($R>1$) settings, and study the impact of distributional shifts between the data sources. Detailed data-generating processes are provided in the Supplementary Material. 


\begin{table}[htb!]
\centering
\small
\setlength{\tabcolsep}{5pt} 
\renewcommand{\arraystretch}{1.2} 
\scalebox{0.9}{
\begin{tabular}{lcccccc}
\toprule
\textbf{Setting} & (I) & (II) & (III) & (IV)  & (V) & (VI) \\
\midrule
Missing Mechanism & MCAR & MCAR & MCAR & CAM  & Misaligned & CAM \\
Number of $R$ & 1 & 1 & 2 & 1,2 & 4 & 1 \\
Response Type & Continuous & Binary & Continuous & Continuous  & Continuous & Continuous \\
$\X_{\Gamma_r}|\X_{\Gamma}$ & Linear & Nonlinear & Linear & Linear & Linear & Linear \\
$Y|\X$ & Linear & Nonlinear & Linear & Nonlinear/Linear & Linear & Linear \\
High-dimensional $\X$ & -- & -- & -- & -- & -- & Yes \\
\bottomrule
\end{tabular}
}
\caption{\label{tab:simu:setting} \textbf{Summary of simulation setups.} 
Each column  (I)–(VI) represents a distinct simulation setting varying by {missing mechanisms (MCAR, CAM, presence of Misaligned sources)}, response type (continuous or binary), number of sources $R$, and model structure. ``Linear/Nonlinear" indicates whether nonlinear effects are included in the generating mechanism or not.}
\end{table}

We summarize methods under comparison as below. {\bf Preliminary}: The preliminary estimator $\widetilde{\boldsymbol{\gamma}}$. {\bf DEFUSE:} $\widehat{\beta}_{\DEF}$ as introduced in Section \ref{sec:method}. For our method, we include different versions of base (anchor) control functions and reweighting schemes. The suffixes {\bf Eg~\ref{example:const}},  {\bf Eg~\ref{example:line}} and {\bf Eg~\ref{example:krr}} indicate constant, linear and KRR reweighting schemes described in Examples \ref{example:const}, \ref{example:line} and \ref{example:krr}. In main paper, we specify  the base control function $\widehat{\bphi}$ to be the AI-predictor of $\c\trans\widetilde{\S}$ using all observed variables in each source $\Lsc\Msc_r$; Results from reduced working model strategy \citeg{zhao2023heterogeneous, wang2023survey} of $\widehat{\bphi}$ and direct utilization of KRR are provided in Supplementary Material.


We also include the following existing approaches. {\bf MICE:} Multiple imputation by chained equations \citep{van2011mice}. {\bf MBI:} The multiple blockwise imputation method \citep{xue2021integrating} combining $\Lsc\Csc$ and $\Lsc\Msc$ data. {\bf HTLGMM:} The data fusion approach \citep{zhao2023heterogeneous} based on the generalized method of moments. {{\bf SC:} the approach \citep{wang2023survey} using {\bf S}ample-reweighting for {\bf C}alibration of the target model. SC and HTLGMM are built upon the same reduced working parametric model of $Y\sim \X_{\Gamma_r}$ have the same asymptotic variance in our MCAR setting.} {In CAM settings, we also include {\bf SC-PW}, a straightforward extension of SC used in \citet{wang2023survey} that leverages $\widehat{h}_{\Lsc\Csc}$ and $\widehat{h}_{\Lsc\Msc_r}$ as propensity weights to adjust for distributional shifts between sources.} {\bf SSB:} The SS and BM fusion method proposed by \citep{song2024semi} using the $\Lsc\Csc$, $\Lsc\Msc$ and $\Usc\Csc$ samples. 
{\bf SSL:} The SS estimator using $\Lsc\Csc$ and $\Usc\Csc$ \citeg{gronsbell2022efficient}. All methods are implemented in the settings when applicable.

In Setting~(I), we assess estimator performance across $\rho\equiv\rho_1 \in {2,3,4,5}$. Table~S2 reports the average relative efficiency (RE) against the preliminary estimator, computed as the inverse of the mean squared error averaged over all target coefficients and the mean absolute bias. Specifically, the preliminary estimator would have the same performance as the GLM estimator using only $\Lsc\Csc$ data when model is correctly specified under MCAR mechanism. All estimators, except MICE, exhibit small biases. Our DEFUSE consistently outperforms other methods in RE. For example, DEFUSE Eg \ref{example:krr} achieves roughly 20\% higher RE than HTLGMM, MBI and SC at $\rho=5$. And RE monotonically increases with more complex reweight method from Eg \ref{example:const} to Eg \ref{example:krr}. 

For Settings~(II) and~(III), we present component-wise RE to preliminary in Table \ref{tab:MCAR:coe RE}, RE under (II) with different $\rho$ in Table S4, and RE with different control function in Table S3. In terms of fusing $\Lsc\Csc$ and $\Lsc\Msc_r$'s, DEFUSE consistently attains higher efficiency than MBI, SSB, HTLGMM, and SC. For example, our method shows about $2$ times higher RE on $\gamma_1$ compared with SSB and MBI with smaller absolute bias, while standard SSL method provides little to no gain when the outcome model is correct. Regardless of whether the outcome model is correctly specified, and regardless of whether there is a single or multiple missing-source, the RE achieved by Eg~\ref{example:krr} moderately outperforms that of the other two examples. For instance, DEFUSE Eg~\ref{example:krr} attains about 25\% higher RE on $\gamma_3$ than Eg~\ref{example:const} in Setting (II) and  13\% higher RE on $\gamma_3$ in Setting (III). 

\begin{table}[htb!]
\centering
\vspace{0.5em}
\renewcommand{\arraystretch}{1.1}
\setlength{\tabcolsep}{7pt}

\scalebox{0.92}{
\begin{tabular*}{\linewidth}{@{\extracolsep{\fill}}lcccccccccc}
\toprule
 & \multicolumn{5}{c}{\textbf{Setting (II)}} & \multicolumn{5}{c}{\textbf{Setting (III)}} \\
\midrule
\textbf{Method} & $\gamma_1$ & $\gamma_2$ & $\gamma_3$ & $\gamma_4$ & $\gamma_5$ &
$\gamma_1$ & $\gamma_2$ & $\gamma_3$ & $\gamma_4$ & $\gamma_5$ \\
\midrule
DEFUSE Eg~\ref{example:const} & 7.05 & 6.92 & 3.76 & 2.71 & 2.90 & 2.13 & 2.13 & 2.12 & 1.43 & 1.57 \\
DEFUSE Eg~\ref{example:line}  & 7.24 & 6.80 & 4.31 & 2.69 & 2.83 & 2.18 & 2.18 & 2.17 & 1.48 & 1.63  \\
DEFUSE Eg~\ref{example:krr}   & 7.24 & 6.89 & 4.71 & 2.77 & 2.92 & 2.41 & 2.40 & 2.39 & 1.57 & 1.81 \\
MICE                       & 4.86 & 10.01 & 4.92 & 0.18 & 0.58 & 0.52 & 0.63 & 0.69 & 0.20 & 0.22 \\
MBI                        & - & - & - &-  & - & 1.06 & 1.09 & 1.06 & 0.95 & 1.01 \\
SSL                        & 1.01 & 1.04 & 1.04 & 0.97 & 1.29 & 1.01 & 1.00 & 1.00 & 0.99 & 1.02 \\
SSB                        & - & - & - & - & - & 1.07 & 1.12 & 1.51 & 0.60 & 0.81 \\
HTLGMM                     & 4.30 & 4.70 & 2.05 & 1.14 & 1.06 &  - & - & - & - & - \\
SC                         & 4.34 & 4.75 & 2.03 & 1.12 & 1.04 &  - & - & - & - & - \\
\bottomrule
\end{tabular*}
}
\vspace{0.3em}
\caption{Relative efficiency (RE) to preliminary on each coefficient in Settings (II) and (III).}
\label{tab:MCAR:coe RE}
\end{table}

In the CAM Setting (IV), we present the Bias, SE and RE with increasing degree of nonlinearity in Table \ref{tab:MAR:single:nonlinear_auxi}, with increasing degree of density shift and $R=2$ in Table S6, with different control functions in Table S5. As shown in Table~\ref{tab:MAR:single:nonlinear_auxi}, the proposed DEFUSE estimators maintain small mean absolute biases relative to their empirical standard errors; across the reported nonlinear settings, the bias-to-SE ratios remain below roughly 1/3 while SC-PW and MBI estimators become increasingly biased as the nonlinearity strengthens with their bias-to-SE ratios approaching 1/2. Also, the RE of DEFUSE consistently outperforms benchmarks and increases with the degree of nonlinearity, rising from approximately $1.77$ to above $1.96$ in average. while SC-PW yields RE $0.98$ and MBI yields RE $0.93$ when nonlinearity $nl=0.8$. At the same time, the RE of DEFUSE generally increases with the degree of nonlinearity, improving from about $1.7$ to nearly $1.9$ while the benchmark MBI and SC-PW is also show RE below one when $nl =0.8$. Setting~(VI) further investigates the CAM scenario with high-dimensional auxiliary features in $\X$, where we vary the number of signal features in ${\Z_r}$. Our design also mimics auxiliary features in EMR studies. The results are summarized in Table~\ref{tab:MAR:single:nonlinear_auxi}. DEFUSE consistently exhibit small absolute biases and achieve substantial efficiency gains compared to the preliminary baseline.

\begin{table}[htb!]
\centering
\vspace{0.5em}
\renewcommand{\arraystretch}{1.1}
\setlength{\tabcolsep}{6pt}

\scalebox{0.75}{
\begin{tabular}{llcccccc}
\toprule
 \textbf{Setting} & & \multicolumn{3}{c}{\textbf{Bias$_{{\rm SE}}$}} 
 & \multicolumn{3}{c}{\textbf{RE}} \\
\midrule
 & \textbf{Degree of Nonlinearity} 
& $nl = 0.4$ & $nl = 0.6$ & $nl = 0.8$ 
& $nl = 0.4$ & $nl = 0.6$ & $nl = 0.8$ \\
\cmidrule(lr){2-8}

\multirow{6}{*}{Setting (IV)} 
 & Preliminary
 & $0.022_{0.112}$ & $0.027_{0.136}$ & $0.033_{0.162}$ 
 & 1.00 & 1.00 & 1.00  \\

 & DEFUSE Eg~\ref{example:const} 
 & $0.024_{0.086}$ & $0.026_{0.102}$ & $0.029_{0.119}$ 
 & 1.74 & 1.89 & 1.97  \\

 & DEFUSE Eg~\ref{example:line} 
 & $0.026_{0.085}$ & $0.032_{0.098}$ & $0.040_{0.114}$ 
 & 1.75 & 1.90 & 1.98  \\

 & DEFUSE Eg~\ref{example:krr} 
 & $0.024_{0.084}$ & $0.031_{0.098}$ & $0.040_{0.114}$ 
 & 1.80 & 1.87 & 1.89  \\

 & MBI 
 & $0.029_{0.082}$ & $0.045_{0.110}$ & $0.061_{0.143}$
 & 1.55 & 1.18 & 0.98  \\

 & SC-PW
 & $0.040_{0.085}$ & $0.061_{0.115}$ & $0.082_{0.147}$
 & 1.45 & 1.10 & 0.93  \\

\toprule

& \textbf{Number of Signal} 
& $s=5$ & $s=10$ & $s=15$
& $s=5$ & $s=10$ & $s=15$ \\
\cmidrule(lr){2-8}

\multirow{4}{*}{Setting (VI)} 
& Preliminary 
& $0.013_{0.090}$ & $0.015_{0.090}$ & $0.012_{0.090}$
& 1.00 & 1.00 & 1.00 \\

& DEFUSE Eg \ref{example:const}
& $0.014_{0.079}$ & $0.015_{0.076}$ & $0.014_{0.077}$ 
& 1.28 & 1.38 & 1.36 \\

& DEFUSE Eg \ref{example:line}
& $0.016_{0.078}$ & $0.018_{0.076}$ & $0.017_{0.076}$ 
& 1.27 & 1.34 & 1.36 \\

& DEFUSE Eg \ref{example:krr}
& $0.016_{0.077}$ & $0.018_{0.075}$ & $0.017_{0.075}$ 
& 1.25 & 1.30 & 1.39 \\

\bottomrule
\end{tabular}
}
\vspace{0.3em}
\caption{Mean absolute bias, empirical standard error, and MSE relative efficiency to the preliminary estimator under Setting (IV) and (VI). Here $nl$ denotes the degree of nonlinearity effect and $s$ denotes the number of signal covariates among the 20 covariates.}
\label{tab:MAR:single:nonlinear_auxi}
\end{table}

\begin{table}[htbp]
\centering
\renewcommand{\arraystretch}{1.1}
\setlength{\tabcolsep}{4.5pt}

\scalebox{0.80}{
\begin{tabular}{lcccccccc}
\toprule
& \multicolumn{2}{c}{\textbf{MD Max}}
& \multicolumn{2}{c}{\textbf{MSD Max}}
& \multicolumn{2}{c}{\textbf{MD Combined}}
& \multicolumn{2}{c}{\textbf{MSD Combined}} \\
\midrule
\textbf{Discrepancy} $\alpha_{\mathrm{mis}}$
& Sensitivity & F1
& Sensitivity & F1
& Sensitivity & F1
& Sensitivity & F1 \\
\midrule
0.2 & 0.01 & 0.68 & 0.03 & 0.77 & 0.01 & 0.67 & 0.07 & 0.85 \\
0.4 & 0.01 & 0.88 & 0.38 & 0.94 & 0.21 & 0.86 & 0.62 & 1.00 \\
0.6 & 0.49 & 0.94 & 0.96 & 1.00 & 0.81 & 0.97 & 1.00 & 1.00 \\
0.8 & 1.00 & 0.99 & 1.00 & 1.00 & 1.00 & 0.99 & 1.00 & 1.00 \\
\bottomrule
\end{tabular}
}
\vspace{0.3em}
\caption{MSD screening performance with increasing discrepancy under Setting (V). ``Sensitivity'' denotes the correct rejection proportion in the single-source experiment, while F1 is computed in the four-source experiment under the true keep pattern $(0,1,0,1)$. ``Max'' is the largest standardized moment over individual basis functions, while ``Combined'' aggregates all basis moments through quadratic form.
}
\label{tab:msd_screening_combined}
\end{table}

In Setting~(V) with misaligned sources violating the CAM assumption, we evaluate the performance of our screening approach based on the mean squared difference (MSD) $\bar\Delta_r^2$ introduced in Section \ref{sec:violation} in Table~\ref{tab:msd_screening_combined}, across varying degrees of distributional shifts and number of misaligned sources. Our performance metrics include F1 score and sensitivity for discerning the misaligned sources from the aligned ones. MSD method outperforms the commonly used mean difference (MD) method discuss in Remark \ref{rem:4} and are more sensitive to the discrepancy. In addition, MSD Combined attains F1 score close to $1$ in all cases while other methods fail to provide consistent selection of the misaligned sources. This result agrees with our theoretical conclusion and discussion in Section \ref{sec:thm:screen}.

\section{Real Examples}\label{sec:real}

\subsection{Alzheimer’s Risk Modeling with NACC data}

We apply DEFUSE to the National Alzheimer’s Coordinating Center (NACC) database \citep{weintraub2018version}, a longitudinal cohort compiled from multiple U.S. Alzheimer’s Disease Research Centers, to build a prognostic model for Alzheimer’s disease (AD). The binary outcome $Y \in \{0,1\}$ indicates cognitive impairment, and covariates $\X$ include demographics (gender, race, age, weight, height), behavioral factors (living situation, alcohol use, smoking), and a clinician-assessed language function score \citep{sonty2003primary}. Eight centers with missing language scores are initially included, among which three are excluded by our screening method in Section~\ref{sec:violation} due to potential misalignment. 

Table~\ref{tab:re:bm:ad} presents the relative efficiency (RE) of DEFUSE (Examples~\ref{example:const}, \ref{example:line}, \ref{example:krr}) compared with the preliminary baseline. DEFUSE with KRR reweighting function (Eg~\ref{example:krr}) consistently outperforms preliminary and DEFUSE Eg~\ref{example:const}, Eg~\ref{example:line}. For example, for the effect of \emph{Alcohol}, DEFUSE Eg~\ref{example:krr} achieves an RE of 2.76, around 35\% higher than its Eg~\ref{example:line} counterpart and 72\% higher than its Eg~\ref{example:const} counterpart. Similar improvements are observed for other fully observed covariates, especially in partially missing \emph{Language} score.

\begin{table}[h]
  \centering
  \scalebox{0.85}{
  \begin{tabular}{|l|c|c|c|c|c|c|c|c|c|}
    \hline \textbf{Method} & \textbf{Sex} & \textbf{Race} & \textbf{Age} & \textbf{Weight} & \textbf{Height} & \textbf{Life} & \textbf{Alcohol} & \textbf{Smoke} & \textbf{Language} \\
    \hline 
    DEFUSE Eg~\ref{example:const} & 1.72 & 1.57 & 2.94 & 2.28 & 1.46 & 2.00 & 1.59 & 1.62 & 1.01 \\
    DEFUSE Eg~\ref{example:line}  & 2.04 & 1.82 & 3.55 & 3.72 & 1.65 & 2.42 & 2.05 & 1.91 & 1.15 \\
    DEFUSE Eg~\ref{example:krr}   & 2.82 & 2.12 & 4.51 & 4.30 & 2.34 & 3.33 & 2.76 & 1.87 & 1.56 \\
    \hline
    \end{tabular}
    }
    \caption{\label{tab:re:bm:ad} Relative efficiency (RE) to the preliminary estimator on NACC data.}
\end{table}

\subsection{Heart Disease Risk Modeling with MIMIC-III data}\label{sec:real:mimic}

We apply DEFUSE to the MIMIC-III dataset \citep{johnson2016mimic} for heart disease (HD) risk modeling, where $Y \in \{0,1\}$ denotes HD status determined by manual chart review \citep{gehrmann2018comparing}. The risk factors $\A$ include age, low high-density lipoproteins (HDL), type II diabetes, and HD-related laboratory markers, while auxiliary features $\X$ are derived from diagnostic and procedure codes. We consider two settings: a single BM ($R=1$) using a composite laboratory risk score (LRS), and a multiple-BM ($R=3$) case involving two partially missing biomarkers—thyroid-stimulating hormone (TSH) and hypochromia (HPO). All BM sources pass the screening in Section~\ref{sec:violation} under $\Z_{r}=\emptyset$, consistent with the MCAR assumption. Thus, we mainly introduce the MCAR results, with the CAM results included in the Supplementary Material.

Table~\ref{tab:combined} reports the RE of the estimators to the preliminary benchmark. DEFUSE with all reweight methods  outperform existing BM data fusion methods. For example, the average RE of DEFUSE Eg \ref{example:krr} is around 80\% higher than those of SC and HTLGMM in the single BM scenario. When our assumption is relaxed to CAM, the RE of DEFUSE to preliminary is still significant and interestingly, the improvement of DEFUSE Eg \ref{example:krr} over Eg \ref{example:const} is more significant compared to the MCAR scenario.

\begin{table}[!h]
    \centering
    \scalebox{0.8}{
    \begin{tabular}{|l|c|c|c|c|c|c|}
        \hline
        \textbf{Method} & \textbf{Age} & \textbf{HDL} & \textbf{Diabetes} & \textbf{LRS} & \textbf{TSH} & \textbf{Hypochromia} \\
        \hline
        \multicolumn{7}{|c|}{\textbf{Single BM Scenario}} \\
        \hline
        DEFUSE Eg~\ref{example:const} & 2.11 & 2.37 & 2.80 & 1.11 & - & - \\
        DEFUSE Eg~\ref{example:line}  & 3.02 & 2.55 & 2.89 & 1.14 & - & - \\
        DEFUSE Eg~\ref{example:krr}   & 3.61 & 2.80 & 3.36 & 1.17 & - & - \\
        HTLGMM & 1.76 & 1.55 &  1.56 & 1.02 & - & - \\
        SC & 1.76 & 1.55 &  1.56 & 1.02 & - & - \\
        \hline
        \multicolumn{7}{|c|}{\textbf{Multiple BM Scenario}} \\
        \hline
        DEFUSE Eg~\ref{example:const} & 1.96 & 2.21 & 2.85 & - & 1.86 & 1.25 \\
        DEFUSE Eg~\ref{example:line}  & 3.28 & 2.43 & 3.36 & - & 3.01 & 1.54 \\
        DEFUSE Eg~\ref{example:krr}   & 3.65 & 3.61 & 5.72 & - & 5.38 & 1.70 \\
        SSL  & 1.00 & 0.99 & 1.06 & - & 0.94 & 1.13 \\
        \hline
    \end{tabular}
    }
    \caption{\label{tab:combined} Relative efficiency (RE) to preliminary estimator in both single BM and multiple BM scenarios on MIMIC-III.}
\end{table}

\section{Discussion}\label{sec:diss}

We develop DEFUSE, a novel approach for robust and efficient data fusion in the presence of blockwise missing covariates, large unlabeled samples, and distributional shifts across data sources. Our approach is data-adaptive and robust to various missing patterns and model misspecification, remains computationally tractable, and accommodates high-dimensional auxiliary features. To tackle source misalignment, we develop a sensitive screening procedure using a debiased MSD measure, which effectively identifies and excludes non-transferable data sources to prevent fusion-induced bias. Crucially, our theoretical developments establish that DEFUSE attains the semiparametric efficiency bound when the candidate class of control functions is sufficiently rich. Furthermore, by integrating flexible nonparametric tools like KRR, our framework provides a computationally tractable mechanism to achieve this theoretical optimality even under complex missing data structures where classical closed-form efficient estimators are unavailable. Theoretical guarantees and extensive experiments confirm DEFUSE's superiority in maintaining robustness and efficiency, offering a practical and powerful solution for modern data fusion problems.

We shall point out several future directions. Our current method requires the $\Lsc\Csc$ sample, which may not be accessible in certain scenarios \citeg{song2024semi}. Note that (\ref{eq:ee:pop}) can actually be identified and constructed with the $\Usc\Csc$ and $\Lsc\Msc_1,\ldots,\Lsc\Msc_R$ samples as long as the union set of the observed covariates in the $\Lsc\Msc_r$’s cover the predictors $\A$. It is interesting to generalize our method to accommodate this setting without $\Lsc\Csc$ data. DEFUSE could be generalized to address high-dimensional regression for $Y\sim \A$. One can use debiased Lasso \citeg{van2014asymptotically} to derive the preliminary estimator and base on it to conduct data fusion. Recent work, such as \citet{kundu2023logistic}, has highlighted the practical need to protect the privacy of individual-level data when fusing datasets collected from different institutions. This need is often addressed by transmitting only summary-level data, e.g., mean vectors and model coefficients \citep{wolfson2010datashield}.

\begingroup
\setstretch{1.0}
\bibliographystyle{apalike} 
\bibliography{library}
\endgroup

\end{document}